\documentclass[onecolumn,manuscript,preprint]{revtex4}

\usepackage{dcolumn}
\usepackage{amsmath}
\usepackage[dvips]{epsfig}      
\makeatletter
\def\btt#1{\texttt{\@backslashchar#1}}%
\DeclareRobustCommand\bblash{\btt{\@backslashchar}}%
\makeatother

\begin{document}
\draft
\title{\bf  Growth model investigation of Vanadium-Benzene Polymer}
\author{X.L. Wang, M.Y. Ni, and Z. Zeng$\footnote{Correspondence author: zzeng@theory.issp.ac.cn}$}
\affiliation {Key Laboratory of Materials Physics, Institute of Solid State Physics, Chinese Academy
of Sciences, Hefei 230031, P.R. China and\\ Graduate School of the Chinese Academy of Sciences }
\date{\today}


\begin{abstract}
Electronic and magnetic properties of V$_n$Bz$_{n+1}$ sandwich clusters are investigated
using density functional theory. Growth model is applied to investigate the property change of
Vanadium-Benzene sandwich clusters. Our results show that, for n$\le$8, all V ions
tend to put their spin in ferromagnetic state, and that the magnetic moments of V$_n$Bz$_{n+1}$
increase linearly with \emph{n}.
Finite-size effects  induce a nonmonotonous behavior of the V$_n$Bz$_{n+1}$ magnetic properties.
 In the case of \emph{n}=8, the electronic properties of V$_n$Bz$_{n+1}$
has the same characteristics with the counterpart of Vanadium-Benzene infinite wire, hence, the critical
length of V$_n$Bz$_{n+1}$ is defined eight. Furthermore,
our results demonstrate that Vanadium-Benzene infinite wire is a proper material for
spin polarized transport and has a high stability in the presence of external electronic and magnetic fields.
\end{abstract}


 \maketitle


\section{\bf Introduction}
Recently, one kind of one-dimensional transition-metal based molecular magnets
have attracted great attentions,\cite{ref000101,ref000102,ref000103,ref0001,ref0002,ref0003,ref0004,ref01008,ref01009} because
these molecular magnets are considered as
potential material used in recording media and spintronics devices.
Among all of them, V$_n$Bz$_{n+1}$ sandwich clusters are a hot object and have a linear structures and ferromagnetic ground states,
those clusters were suggested to be a proper material for spintronics devices, especially for
spin filter and spin-polarized transportation.
Experimental
studies suggested that, for n$\leq$4, V$_n$Bz$_{n+1}$ sandwich clusters have a
one-dimensional structure and possess the magnetic moments increasing monotonously
with \emph{n}\cite{ref0003}. Theoretical investigations on V$_n$Bz$_{n+1}$ provided farther information for those
structural and magnetic properties\cite{ref0004}.
 Maslyuk et al.\cite{ref0005} found that V$_3$Bz$_4$ placed
between magnetic Ni or Co electrodes will act as
almost perfect spin filter, and that the spin polarization of the current can
reach 90\%. Koleini et al.\cite{ref0006} also used spin density functional theory and nonequilibrium Green's function
to investigate the transport properties of V$_n$Bz$_{n+1}$ placed between graphene
or single-wall carbon-nanotube, their results showed that spin polarization of the current could
reach as high as 99\%, and that the polarization increased with cluster size increased.
In the case of Vanadium-Benzene infinite wire (V(Bz)$_{\infty}$), a spin polarized density of states
around Fermi Surface was found by Xiang et al.\cite{ref0007}, thus V(Bz)$_{\infty}$ was deemed to
a perfect material for spin-polarized current transport.

As mentioned above, V$_n$Bz$_{n+1}$ and V(Bz)$_{\infty}$ may be an ideal material for
recording and spin-polarized transport. However, it is very important to find out
the link between V$_n$Bz$_{n+1}$ and V(Bz)$_{\infty}$, in other words, critical length
in which the magnetic clusters have the same properties as V(Bz)$_{\infty}$ is a key
parameter for practical applications and theoretical studies. Additionally, since V$_n$Bz$_{n+1}$
and V(Bz)$_{\infty}$ are expected to be used in recording and spin-polarized transport, they
 will inevitably endure the interference of external electronic and magnetic fields, and
 the stability of V(Bz)$_{\infty}$ in the presence
of external electronic and magnetic fields is also an interesting issue. Hence, in this work, we will mainly discuss
the critical length of V$_n$Bz$_{n+1}$ using growth model
 and the stability of V(Bz)$_{\infty}$ in the presence of external electronic and
magnetic fields.

This paper is organized as follows: the computational method and theoretical model are briefly described
in Sec. II; the results and discussions are presented in Sec. III; and a brief summary is given in Sec. IV.

 \section{Computational Details}
   The linear sandwich of V$_n$Bz$_{n+1}$ is constructed as the following. Firstly, the most stable structure of
   VBz$_2$ is obtained by performing the full relaxation.
   For the ground state of V$_2$Bz$_3$, VBz is added to the stable  VBz$_2$
   with the added Bz rotating 0$^\circ$,15$^\circ$ and 30$^\circ$, respectively, furthermore, the magnetic moment of the added V is placed
   in the direction of up and down. Therefore, on the base of the optimized VBz$_2$, six different initial
   structures, which are fully relaxed without any symmetry constraint, are used for searching the ground state of V$_2$Bz$_3$.
   Following the same way,
   the completely relaxed ground states of V$_n$Bz$_{n+1}$ are obtained from the optimized V$_{n-1}$Bz$_{n}$. This process
    is named as the growth model which is widely used in cluster science.
 As far as we know, it is the first time that growth model is applied on this kind
of molecular magnets. In our opinion, the growth model can simulate the growth process of V$_n$Bz$_{n+1}$
   more accurately.

The properties of V$_n$Bz$_{n+1}$ and V(Bz)$_{\infty}$ are calculated
 based on the density functional theory,\cite{ref0012}, and the generalized gradient
 approximation (GGA) with the Perdew, Burke and Ernzerhof functional\cite{ref0013} is used.
  Valence electrons are expanded by linear combination of numerical atomic orbital formalism
   as implemented in SIESTA code\cite{ref0009,ref0010,ref0011},
  and core electrons are described by the Troullier-Martins pseudopotentials.\cite{ref0014}
We use double-$\zeta$ basis (DZ) for
   C and H orbitals and a double-$\zeta$ basis plus polarization (DZP) for the transition metal
   orbitals.
  We have fully tested them on the properties
   of benzene, methane and bulk V, in which the results agree well with
   experimental values or full-electron calculations.

   All the atomic positions
are fully relaxed until forces are less than 0.04 eV/\AA\space and the convergence criterion for density matrix
is 10$^{-4}$.

\section{\bf Results and Discussions}

The optimized sandwich VBz$_2$ with \emph{D$_{6h}$} symmetry has an integer magnetic moment of 1$\mu_{B}$.
V has a localized magnetic moment (1.23$\mu_{B}$), and benzene are anti-parallelly polarized
with the magnetic moment of 0.115$\mu_{B}$ distributed over 6 C atoms.
Also, V$_2$Bz$_3$ has the \emph{D$_{6h}$} symmetry, and the rotated states with the edge Bz rotating
15$^\circ$ and 30$^\circ$ are semi-stable configurations. These structural characteristics are maintained
until \emph{n} increases to 8.
 As shown in Table I, the total energies of
V$_n$Bz$_{n+1}$ increase with the rotation angle of the edge Bz increasing. This  phenomenon also appears
in V(Bz)$_{\infty}$.\cite{ref0005}

It is one of our purposes to find
the link between V$_n$Bz$_{n+1}$ and V(Bz)$_{\infty}$ for practical applications and theoretical studies.
We will use the growth model described above to  search the critical length of V$_n$Bz$_{n+1}$ by inspecting the electronic properties of
 V$_n$Bz$_{n+1}$ and V(Bz)$_{\infty}$.
 Fig. 1 presents the evolution of the density of states (DOS) of V$_n$Bz$_{n+1}$ from $n=1$ to $n=\infty$.
  When the number of \emph{n} closes to 8, the DOS of V$_n$Bz$_{n+1}$ has similar characteristics
 with the counterpart of V(Bz)$_{\infty}$, especially around Fermi Level. Hence, eight layer is the critical length
 for V$_n$Bz$_{n+1}$. After \emph{n}$\geq$8, the properties of V$_n$Bz$_{n+1}$ do not change remarkably.
Note that, in the cases of V$_n$Bz$_{n+1}$, there are additional states distributed from -5 eV to -4 eV compared with the DOS of
V(Bz)$_{\infty}$.
 These isolated states are from the edge benzenes. Because there are two edge benzenes for
any one of V$_n$Bz$_{n+1}$, the magnitude and shape of the
isolated states do not change as \emph{n} changes from 1 to 8.
 In order to further confirm the critical length is 8, the properties of V$_9$Bz$_{10}$ are investigated, the DOS of
V$_9$Bz$_{10}$ is shown in the inset (a) of Fig 2, we can found that, around the Fermi Level, the DOS of V$_9$Bz$_{10}$ changes
very weakly comparing with the counterpart of
V$_8$Bz$_{9}$ and has similar characteristics with the DOS of V(Bz)$_{\infty}$.
The DOS of V(Bz)$_{\infty}$
around Fermi Surface shown in Fig. 1(h) is polarized and thus  V(Bz)$_{\infty}$
is expected to be a proper material for spin-polarized transport.

The total energy calculations exhibit that all V$_n$Bz$_{n+1}$ have ferromagnetic ground states, in other words,
the added V ion intends to put his magnetic moment parallel with other V ions.
 As seen in Fig. (2), the magnetic moments of
V$_n$Bz$_{n+1}$ increase linearly with the number of V ions increasing,
which is consistent with the experimental results (\emph{n}$\leq$3)\cite{ref0003} and
 theoretical investigation for \emph{n}$\leq$6.\cite{ref0004}.
 However,
 the added V ion has an anti-parallel magnetic moment with the other V ions named as anti-ferromagnetic states is a second
stable state. The energy difference, which can be deemed to the energy costing(ES) to inverse the magnetic
moment of edge V ion, between the second stable and ground state are shown in Table II. An nonmonotonous behavior
 of ES is shown in Table II, from \emph{n}=2 to \emph{n}=4,
 ES increases with \emph{n} in accord with Weng's results.\cite{ref00141} Whereas, ES decreases from 102 meV to
 67 meV while \emph{n} changing from 4 to 5. For \emph{n}$\leq$8, the drastic variations of ES due to
 finite-size effects are found. In order to confirm this phenomenon, we do not take into account the structural relaxation, but only the
 magnetic interaction between the edge V ion and the other V ions are considered. Hence, all atomic positions are fixed and
 V-Benzene distance is fixed as 1.70\AA. The data appearing
 in the no-relaxation row of Table II  also clearly present the nonmonotonous behavior.
 The same phenomenon were found in magnetic anisotropy of one-dimensional nanostructures of transition metals\cite{ref0015,ref0016},
where D\`{a}vila et al.\cite{ref0015} reported strong oscillations of the magnetic anisotropy energy as the length of the
one-dimensional chains.

As mentioned above, V$_n$Bz$_{n+1}$ and V(Bz)$_{\infty}$ are seem to be an deal materials for
recording and spin-polarized transport. While V$_n$Bz$_{n+1}$ and V(Bz)$_{\infty}$
are used for spintronics devices, they will be  exposed to additional electronic and magnetic fields.
External fields can influence the properties of magnetic materials strongly,
for example, Son et al.\cite{ref0017} found that the magnetic properties of graphene nanoribbons can
be controlled by the in-plane external electric fields applied through the zigzag-shaped
edges.
Also, in some magnetic systems, heat, pressure and magnetic field\cite{ref0018} can induce
the spin-crossover phenomenon which originates from the transition between the different spin states.
Spin-crossover complexes include organo-metals\cite{ref0019} and diluted magnetic semiconductors.\cite{ref0020}
 The main characteristics of spin-crossover complexes is multi-spin states,
hence, the total energies of V(Bz)$_{\infty}$ unit cell are calculated with the magnetic moment fixed. The calculated total
energies \emph{vs} total magnetic moment shown in Fig. (3) implicit that  there are only
one minimum located at 1\emph{$\mu$}$_B$ for V(Bz)$_{\infty}$, which means that multi-spin states do not
exist in V(Bz)$_{\infty}$. Therefore V(Bz)$_{\infty}$ will be a stable material in the presence of
magnetic field. Furthermore, we apply a transverse electronic field paralleled to benzene molecule V(Bz)$_{\infty}$, and tune the magnitude of electronic
field from 0.0 eV/Ang to 0.5 eV/Ang. The DOSs of V(Bz)$_{\infty}$
in the presence of 0.0 ev/Ang and 0.5 ev/Ang electronic field are shown in Fig. (4).
The electronic structure and magnetic moment of V(Bz)$_{\infty}$ almost are not changed in the presence of
external electronic field. In a word, V(Bz)$_{\infty}$ has good stability in the
 presence of electronic and magnetic field.

  \section{\bf SUMMARY}
We have systematically investigated the structural, electronic and magnetic properties
of V$_n$Bz$_{n+1}$ (\emph{n}$\leq$8). In the case of \emph{n}=8, the electronic
structure of V$_n$Bz$_{n+1}$ has the same characteristics with the counterpart of V(Bz)$_{\infty}$,
hence, eighth-layer is the critical layer of V$_n$Bz$_{n+1}$. In these
 V$_n$Bz$_{n+1}$ system, V ions are arranged in ferromagnetic state. Meaningfully,
 our results reveal an interesting nonmonotonous magnetic behavior caused by finite-size effect, and the energy cost for
 reversing the magnetic moment of the edge V oscillates with \emph{n}.
 V(Bz)$_{\infty}$ is a proper material for spin-polarized transport and has high stability in the
 presence of electronic and magnetic field.

  \section{\bf Acknowledgements}

This work was supported by the National Science Foundation of China
under Grant No 10774148 and No 10847162, the special Funds for Major
State Basic Research Project of China(973), Knowledge Innovation Program of Chinese Academy of
Sciences, and  Director Grants of CASHIPS. Part of the calculations
were performed in Center for Computational Science of CASHIPS and
the Shanghai Supercomputer Center.

\newpage
\begin{table}[tb]
\caption{ The calculated total energy (relative to the lowest energy configuration of 0$^{\circ}$ rotation)
 of V$_n$Bz$_{n+1}$ with the new
added benzene rotating 15$^{\circ}$ and 30$^{\circ}$.}

\begin{ruledtabular}
\begin{tabular}{cccccccc}
  meV  &V$_2$Bz$_3$  &V$_3$Bz$_4$   &V$_4$Bz$_5$    &V$_5$Bz$_6$  &V$_6$Bz$_7$  &V$_7$Bz$_8$  &V$_8$Bz$_9$ \\
\hline
  15$^{\circ}$        &10   &20  &81  &35   &61   &12   &12\\
  30$^{\circ}$        &25   &44  &95  &69   &92   &25   &36\\
\end{tabular}
\end{ruledtabular}
\end{table}
\clearpage

\newpage
\begin{table}[tb]
\caption{ The energy costing (ES) to inverse the magnetic
moment of edge V ion in the cases of growth-model and no-relaxation, respectively.}

\begin{ruledtabular}
\begin{tabular}{cccccccc}
  meV  &V$_2$Bz$_3$  &V$_3$Bz$_4$   &V$_4$Bz$_5$    &V$_5$Bz$_6$  &V$_6$Bz$_7$  &V$_7$Bz$_8$  &V$_8$Bz$_9$ \\
\hline
  growth-model        &1   &8  &102  &67   &73   &60   &47\\
  no-relaxation        &10   &3  &90    &120   &84    &69    &51\\
\end{tabular}
\end{ruledtabular}
\end{table}
\clearpage

\begin{figure}[tp]
\vglue 1.0cm
\newpage
\caption{\label{fig:epsart} The total DOS of V$_n$Bz$_{n+1}$ (n$\leq$8) and V(Bz)$_{\infty}$.
The vertical dashed line represents the Fermi level.}
\caption {Total magnetic moments of V$_n$Bz$_{n+1}$ as a function of \emph{n},
and the DOS of V$_9$Bz$_{10}$ and views of V$_7$Bz$_8$ are shown in(a) and (b) of the inset, respectively.}


\caption {The total energy \emph{vs} the magnetic moment of V(Bz)$_{\infty}$.}


\caption
{The total DOS of V$_n$Bz$_{n+1}$ in the presence of external electronic field
, (a) and (b) for the electronic field are 0.0 eV/Ang and 0.5 eV/Ang,
 respectively.}

\end{figure}
\clearpage

\newpage
\begin{figure*}[htbp]
\center
{$\Huge\textbf{Fig. 1  \underline{Wang}.eps}$}
\vglue 3.0cm
\includegraphics[width=17cm,height=13cm,angle=360]{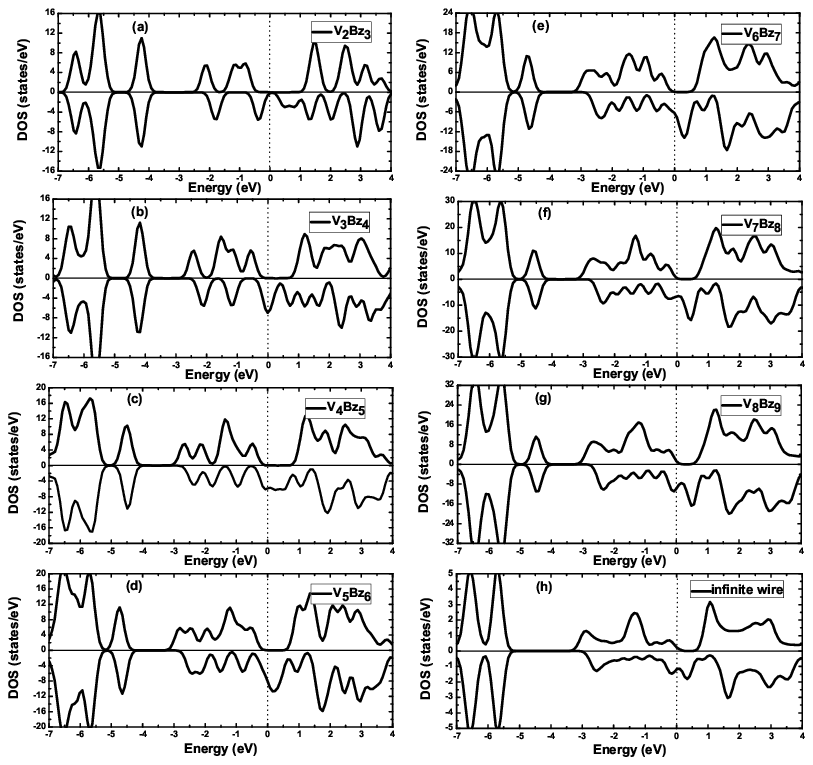}
\end{figure*}
\clearpage
\newpage
\begin{figure*}[htbp]
\center
{$\Huge\textbf{Fig. 2  \underline{Wang}.eps}$}
\vglue 3.0cm
\includegraphics[width=14cm,height=10cm,angle=360]{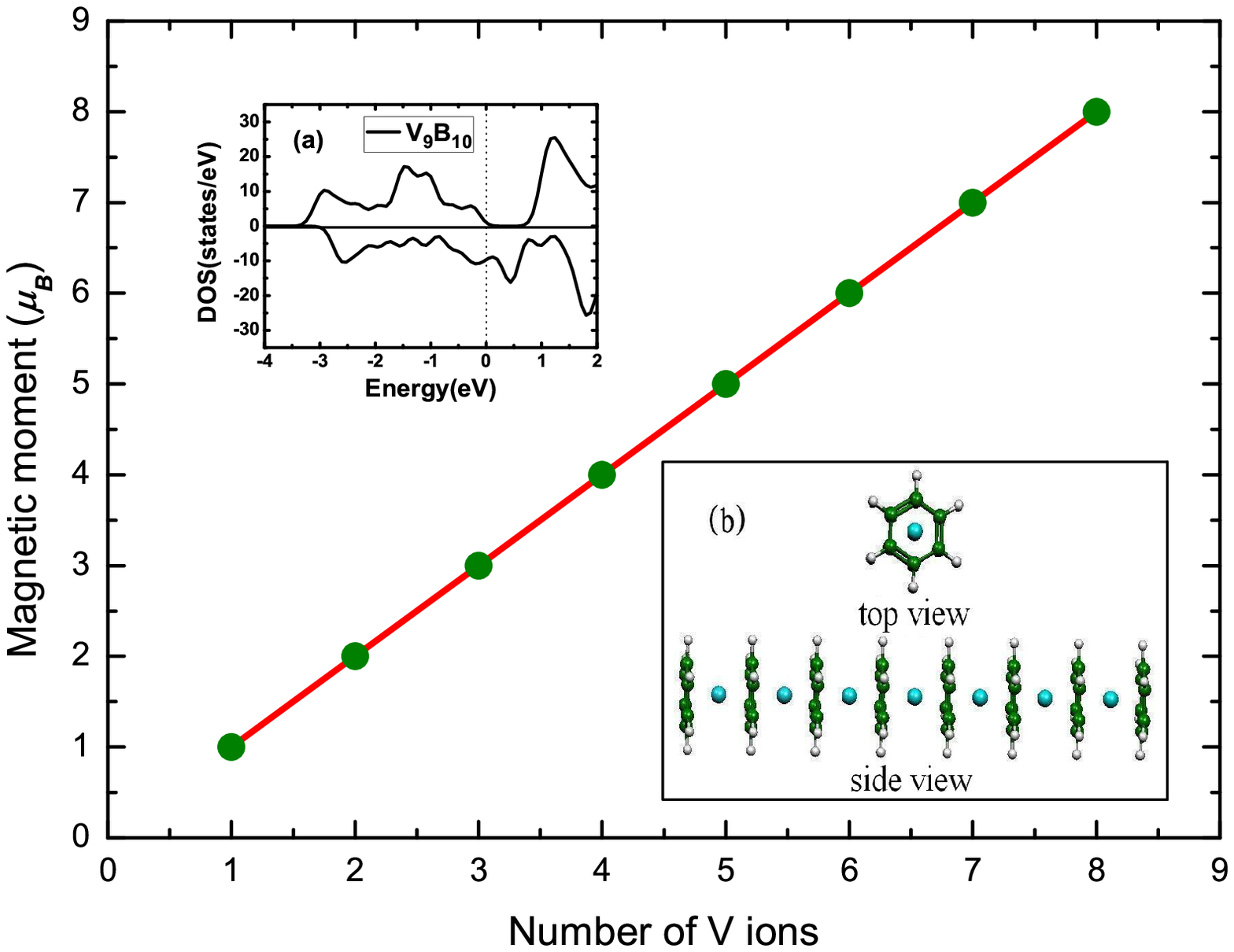}
\end{figure*}
\clearpage
\newpage
\begin{figure*}[htbp]
\center
{$\Huge\textbf{Fig. 3  \underline{Wang}.eps}$}
\vglue 3.0cm
\includegraphics[width=10cm,height=9cm,angle=360]{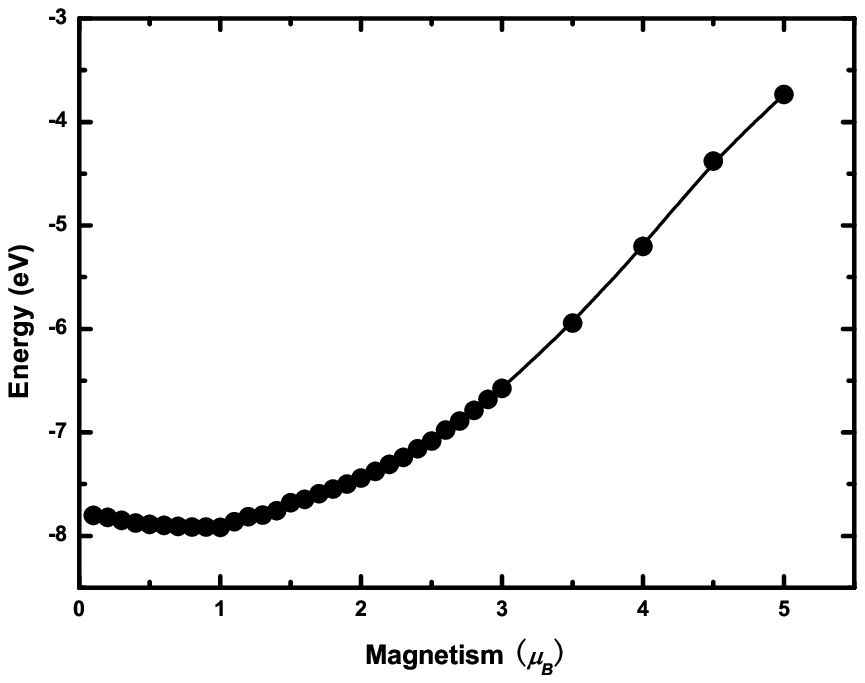}
\end{figure*}
\clearpage
\newpage
\begin{figure*}[htbp]
\center {$\Huge\textbf{Fig. 4  \underline{Wang}.eps}$}
\vglue 3.0cm
\includegraphics[scale=2.0,angle=360]{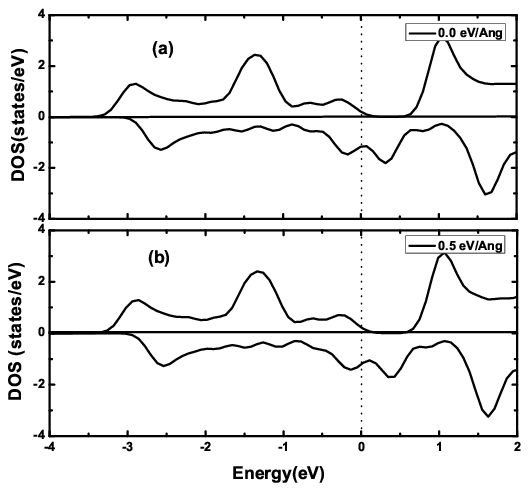}
\end{figure*}
\clearpage
\end{document}